\documentclass[12pt]{article}

\usepackage{epsfig}

\def  \bcen   {\begin{center}}
\def  \ecen   {\end{center}}
\def  \beq    {\begin{equation}}
\def  \eeq    {\end{equation}}
\def  \beqa   {\begin{eqnarray}}
\def  \eeqa   {\end{eqnarray}}
\def  \nn     {\nonumber}

\def  \btokrhokll {B \to K_1 (\to \rho K) \ell^+ \ell^- } 

\begin{document} 

\renewcommand{\thefootnote}{\fnsymbol{footnote}} 
 
\thispagestyle{empty}

\begin{flushright}
KEK-TH-1148, \\
LYCEN 2007-03.
\end{flushright}
\vskip 2cm
\bcen
{\bf \Large \boldmath $B \to K_1(1270) \left(\to \rho K \right) \ell^+ \ell^-$ in LEET}     
\vskip 2cm
{\large \sc S. Rai Choudhury$^1$\footnote{src@physics.du.ac.in}, 
A. S. Cornell$^2$\footnote{cornell@ipnl.in2p3.fr}, 
Naveen Gaur$^3$\footnote{naveen@post.kek.jp}}
\vskip 1cm
{\sl $^1$ Center for Theoretical Physics, Jamia Millia University, Delhi, India. \\
$^2$ Universit\'e de Lyon 1, Institut de Physique Nucl\'eaire, Villeurbanne, France.\\
$^3$ Theory Division, KEK, 1-1 Oho, Tsukuba, Ibaraki 305-0801,Japan. } 
\ecen

\vskip 1.5cm
 \begin{abstract}
Flavour Changing Neutral Current decays of the $B$-meson are a very
useful tool for studying possible physics scenarios beyond the
Standard Model (SM), where of the many FCNC modes radiative, purely
leptonic and semi-leptonic decays of the $B$-meson are relatively
clean tests. Within this context the BELLE collaboration has searched
for the $B \to K_1(1270) \gamma$ process and provided an upper bound on
this decay. In this work we have used this upper bound in studying the
angular coorelations for the related semi-leptonic decay mode $B \to
K_1(1270) (\to \rho K) \ell^+ \ell^-$, where we have used the form
factors that have already been estimated for the $B \to K_1(1270) \gamma$
mode. Note that the additional form factors that are required were
calculated using the Large Energy Effective Theory (LEET).  
\end{abstract}

\vfill \eject 


\section{Introduction \label{sec:intro}} 

\par The Flavour Changing Neutral Current (FCNC) decays of the
$B$-meson are an important tool for investigating possible physics
scenarios beyond the Standard Model (SM), where such decays are
forbidden at tree level. It is for this reason that FCNC processes are
very sensitive to possible small corrections that may be a result of
any modification to the SM, or from some new interactions. Of the FCNC
decays the radiative mode $B \to K^* \gamma$ has been experimentally
measured, with a lot of theoretical work also having gone into its
study. A related decay, $B \to K^* \ell^+ \ell^-$, has also been
observed experimentally. This latter process offers many more
observables for confrontation with theory (such as Forward-Backward
(FB) asymmetries, polarizations and angular correlations between the
final state particles etc.), where the theoretical work for this
subject has spawned many investigations. Recently the radiative mode
$B \rightarrow  K^*_2  \gamma$ has also been observed with good limits
also being set on the modes $B \rightarrow K_1(1270,1400) \gamma$,
where the $K_1$'s are the $1^+$ resonances. The numbers for these
rates are comparable to those for the $K^*(890)$ resonance case 
and we may expect that with more data becoming available the related
$B \to K_1 \ell^+ \ell^-$ would be observed just as with the $K^*(890)$
case. 
Furthermore, as with the $K^*(890)$, such data will provide an
independent opportunity to test the predictions of the SM.     

\par In this paper we study the angular distribution of the rare
$B$-decay $B \to K_1(1270) \left(\to \rho K \right) \ell^+ \ell^-$,
which may be expected to be observed in future B-factories. We use the
standard effective Hamiltonian approach, and use the form factors that
have already been estimated for the corresponding radiative decay $B
\to K_1(1270) \gamma$. The additional form factors for the dileptonic
channel are estimated using the Large Energy Effective Theory (LEET),
which enables one to relate the additional form factors to the form
factors of the radiative mode. Our results provide, just like in the
case of the $K^*(890)$ resonance, an opportunity for a straightforward
comparison of the basic theory with experimental results, which may be
expected in the near future for this channel. Recall that the physical
$K_1(1270)$ and $K_1(1400)$ states are the mixture of $K_{1A}$
($^3P_1$ state) and $K_{1B}$ ($^1P_1$ state) states
\cite{Cheng:2004yj}: 
\beqa
K_1(1270) &=&  K_{1A} sin\theta + K_{1B} cos\theta \,\,\, , \nonumber
\\ 
K_1(1400) &=&  K_{1A} cos\theta - K_{1B} sin\theta \,\,\, ,
\label{eq:1}
\eeqa
where $\theta$ is the $K_1(1270) - K_1(1400)$ mixing angle. Cheng {\sl
  et al.}\cite{Cheng:2004yj} proposed two possible solutions for this
angle, namely $\theta = \pm 37^\circ$, $\pm 58^\circ$. Of these
possibilities the negative values of the mixing angles predict the
Branching Ratio (BR) of $B \to K_1(1400) \gamma$ to be more than that
of $B \to K_1 (1270) \gamma$, which is disfavoured from experimental
data (although not ruled out). For our work we have taken the positive
values of the mixing angles.  

\par The paper is organized as follows: In section II we
will give the relevant effective Hamiltonian and the LEET form factors
for the process under consideration. In section III we will give the
expressions for the differential decay rate for the semi-leptonic
decay mode under consideration. Finally we will conlude with our
results in section IV.  


\section{The Effective Hamiltonian and form factors \label{section2}}

\par The short distance contribution to the decay $B \to K_1 \ell^+ \ell^-$ is governed by the quark level decay $b\to s \ell^+\ell^-$, and where the SM operator basis can be described by the effective Hamiltonian:
\beqa
{\cal H}^{eff} &=& \frac{G_F \alpha}{\sqrt{2} \pi} V_{ts}^* V_{tb} \Bigg[
- 2 C_7^{eff} m_b \left(\bar{s}_R i \sigma_{\mu \nu} \frac{q^\nu}{q^2}
  b_R \right) \left(\bar{\ell} \gamma_\mu \ell \right)
+ C_9^{eff} \left(\bar{s}_L \gamma_\mu b_L \right) 
          \left(\bar{\ell} \gamma^\mu \ell \right)    \nonumber \\
&& \hspace{2cm} + C_{10} \left(\bar{s}_L \gamma_\mu b_L \right) 
          \left(\bar{\ell} \gamma^\mu \gamma_5 \ell \right)
\Bigg] \,\,\, . \label{eq:2}
\eeqa
We can rewrite the above effective Hamiltonian in the following form:
\beqa
{\cal H}^{eff} &=& \frac{G_F \alpha}{\sqrt{2} \pi} V_{ts}^* V_{tb} \Bigg[
- 2 C_7^{eff} m_b \left(\bar{s}_R i \sigma_{\mu \nu} \frac{q^\nu}{q^2}
  b_R \right) \left(\bar{\ell} \gamma_\mu \ell \right)
+ \left( C_9^{eff} - C_{10} \right)  \left(\bar{s}_L \gamma_\mu b_L \right) 
          \left(\bar{\ell}_L \gamma^\mu \ell_L \right)  \nonumber \\
&& \hspace{2cm} + \left( C_9^{eff} + C_{10} \right)
 \left(\bar{s}_L \gamma_\mu b_L \right) 
          \left(\bar{\ell}_R \gamma^\mu \ell_R \right) \Bigg] \,\,\, , \nonumber\\
 &=& \frac{G_F \alpha}{\sqrt{2} \pi} V_{ts}^* V_{tb} 
\Bigg[ - 2 i C_7^{eff} m_b \frac{q^\nu}{q^2} \left(T_{\mu \nu} + T^5_{\mu \nu}\right)   
 \left(\bar{\ell} \gamma^\mu \ell \right) + \left( C_9^{eff} - C_{10} \right) (V - A)_\mu 
   \left(\bar{\ell}_L \gamma^\mu \ell_L \right)   \nonumber \\
&& \hspace{2cm} + \left( C_9^{eff} + C_{10} \right) (V - A)_\mu 
   \left(\bar{\ell}_R \gamma^\mu \ell_R \right) \Bigg] \,\,\, , \label{eq:3}
\eeqa
with 
\beqa
V_\mu &=& {1 \over 2} \left(\bar{s} \gamma_\mu b\right) \,\,\, , \label{eq:4} \\
A_\mu &=& {1 \over 2} \left(\bar{s} \gamma_\mu \gamma_5 b\right) \,\,\, , \label{eq:5} \\
T_{\mu\nu} &=& {1 \over 2} \left( \bar{s} \sigma_{\mu \nu} b \right) \,\,\, , \label{eq:6}  \\
T^5_{\mu\nu} &=& {1 \over 2} \left( \bar{s} \sigma_{\mu \nu} \gamma_
  5b \right) \,\,\, . \label{eq:7}  
\eeqa
In equation (\ref{eq:3}) we have used the $(V-A)$ structure for the hadronic part (except for $C_7$). Note that this structure doesn't change under the transformation $V \leftrightarrow - A$ and $T_{\mu \nu} \leftrightarrow T^5_{\mu \nu}$. Furthermore, we can relate the hadronic factors of $T_{\mu\nu}$ and $T^5_{\mu \nu}$ by using the identity\footnote{Where we have used the convention that $\gamma_5= i \gamma^0 \gamma^1 \gamma^2 \gamma^3$ and that $\varepsilon_{0123}=1$.}:
$$
\sigma_{\mu \nu} = - \frac{i}{2} \varepsilon^{\mu \nu \rho \delta}
\sigma_{\rho \delta} \gamma_5 \,\,\, .
$$

\par In this work we shall closely follow the notation used by Kim {\it et al.}\cite{Kim:2000dq} by defining the form factors of $K_1(1270)$ as:  
\beqa
\langle K_1 (p') | \bar{s} \gamma_\mu b | B (p) \rangle &=& 
- f \epsilon_\mu^* - a_+ (\epsilon^*.p) (p + p')_\mu 
- a_- (\epsilon^*.p) (p - p')_\mu \,\,\, , \label{eq:8} \\
\langle K_1 (p') | \bar{s} \gamma_\mu \gamma_5 b | B (p) \rangle &=& 
- i g \varepsilon_{\mu\nu\lambda\sigma} \epsilon^{* \nu} (p + p')^\lambda
(p - p')^\sigma \nonumber \\
&=& 2 i g \varepsilon_{\mu\nu\lambda\sigma} \epsilon^{* \nu} p^\lambda
     (p')^\sigma \label{eq:9} \,\,\, , \\
\langle K_1 (p') | \bar{s} \sigma_{\mu\nu} \gamma_ 5 b | B (p) \rangle &=& 
g_+ \varepsilon_{\mu\nu\lambda\sigma} \epsilon^{* \lambda} (p + p')^\sigma
+ g_- \varepsilon_{\mu\nu\lambda\sigma} \epsilon^{* \lambda} (p - p')^\sigma \nonumber \\
&& \hspace{1.5cm} + h \varepsilon_{\mu\nu\lambda\sigma} (p + p')^\lambda (p - p')^\sigma 
(\epsilon.p) \,\,\, , \label{eq:10}   \\
\langle K_1 (p') | \bar{s} \sigma_{\mu\nu} b | B (p) \rangle &=& 
- i g_+ \left[ \epsilon^*_\nu \left(p + p'\right)_\mu 
       - \epsilon_\mu^* \left(p + p'\right)_\nu \right]
- i g_- \left[ \epsilon^*_\nu \left(p - p'\right)_\mu 
       - \epsilon_\mu^* \left(p - p'\right)_\nu \right] \nonumber \\
&& \hspace{1.5cm} - i 2 h \left( p_\mu p'_\nu - p_\nu p'_\mu \right)
\left(\epsilon^*.p\right) \,\,\, . \label{eq:11}   
\eeqa
From the above equations we can observe that there are seven form factors which govern the $B \to K_1$ transition, where we will now relate these form factors using the LEET approach. Note that the advantage of writing the form factors in this form is that the expressions of the amplitudes (${\cal A}_R, {\cal A}_L$) are as given in equations (17) and (18) in Kim {\it et al.}\cite{Kim:2000dq}, due to the symmetry of the expressions under the exchange $V \leftrightarrow -A$, $T_{\mu \nu} \rightarrow T^5_{\mu \nu}$, $T^5_{\mu \nu} \rightarrow T_{\mu \nu}$.

\par Note also that Cheng and Chua have parameterized the tensorial form factors for the $B \to K_1$ transition as \cite{Cheng:2004yj}:  
\beqa
\langle K_{1A,1B}(p') | \bar{s} i \sigma_{\mu\nu} q^\nu \left(1 +
\gamma_5\right) b | B(p) \rangle 
&=& i \varepsilon_{\mu\nu\lambda\rho} \epsilon^{*\nu} P^\lambda q^\rho
Y_{A1,B1} + \left(\epsilon^*_\mu P.q - P_\mu \epsilon^*.q \right) Y_{A2,B2}
         \nonumber \\ 
&& \hspace{1.5cm} + \epsilon^*.q \left[ q_\mu - P_\mu \frac{q^2}{P.q} \right] Y_{A3,B3} \,\,\, ,
\label{eq:12}
\eeqa
where $P = p + p'$ and $q = p - p'$. The $K_{1A}$ and $K_{1B}$ states are the angular momentum eigenstates as defined in equation (\ref{eq:1}). Using equation (\ref{eq:1}) we can define the physical $B \to K_1(1270)$ form factors as: 
\beq
Y_i^{B \to K_1(1270)} = Y_{Ai}(q^2) sin\theta + Y_{Bi}(q^2) cos\theta \,\,\, ,
\quad i = 1, \,\,\, 2, \,\,\, 3 \,\,\, .
\label{eq:13}
\eeq
The parameterizations of the form-factors $Y_i$, as used by Cheng and Chua, are given in Appendix \ref{appendix:a}. These form factors can be related to the tensorial form factors given in equations (\ref{eq:9})-(\ref{eq:12}) by:  
\beqa
Y_{1}^{B \to K_1(1270)} &=& - g_+ \,\,\,  , \label{eq:14} \\
Y_{2}^{B \to K_1(1270)} &=& - g_+ - g_- \frac{q^2}{P.q} \,\,\,
, \label{eq:15} \\ 
Y_{3}^{B \to K_1(1270)} &=& g_- + h (P.q) \,\,\, . \label{eq:16} 
\eeqa

\par Using the LEET approach as given by Charles {\it et al.}\cite{Charles:1998dr} we can obtain the following relations between the form factors: 
\beqa
f &=& 2 M E g \,\,\, , \label{eq:17} \\
g_+ &=& - g M \,\,\, , \label{eq:18}  \\
g_- &=& g M  \,\,\, , \label{eq:19}  \\
a_+ &=& - a_- = - (g + h M) \,\,\, , \label{eq:20} 
\eeqa
where $M$ is the mass of the parent hadron and $E$ is the energy of the daughter hadron. We now define all the form factors in terms of just two independent form factors ($g$ and $h$).  

\par Using the LEET relations as given in equations (\ref{eq:17})-(\ref{eq:20}) and equations (\ref{eq:13})-(\ref{eq:16}), the form factors for the $B \to K_1$ transition can be related to Cheng and Chua's form factors as\footnote{It is important to note that the notations of the Levi-Civita tensor in Cheng and Chua's paper \cite{Cheng:2004yj} differs from the notation of Kim {\it et al.} \cite{Kim:2000dq} by a overall negative sign. We are following the notation of Kim {\it et al.}}:
\beqa
g_+ &=& - Y_{1}^{B \to K_1(1270)} \,\,\, , \nonumber \\
g_- &=&   Y_{1}^{B \to K_1(1270)} \,\,\, , \nonumber \\
g   &=&   \frac{Y_1^{B \to K_1(1270)}}{M} \,\,\, , \nonumber \\
f   &=&   2 E Y_1^{B \to K_1(1270)} \,\,\, , \nonumber \\
h   &=&   \frac{Y_3^{B \to K_1(1270)} - Y_1^{B \to K_1(1270)}}{M^2 -
  m_V^2} \,\,\, , \nonumber \\ 
a_+ &=& - a_- = \frac{Y_1^{B \to K_1(1270)} m_V^2 - Y_3^{B \to
    K_1(1270)} M^2}{M \left(M^2 - m_V^2\right)} \,\,\, , \label{eq:21}
\eeqa
where $M$ is the mass of the $B$-meson and $m_V$ is the mass of the $K_1$. 


\section{Kinematics and differential decay rate \label{section:3}}

\par In the following it is convenient to define our kinematics in terms of the following vectors:
$$P = p' = p_\rho + p_K \,\,\, , \quad  
Q = p_\rho - p_K \,\,\, , \quad 
L = p_+ + p_- \,\,\, , \quad 
N = p_+ - p_- \,\,\, .$$

\par Subsequently the decay mode $K_1 \to \rho K$, of the $K_1$ meson, can be parameterized by the matrix element \cite{Roca:2003uk}:
\beqa
{\cal M}\left(K_1 (p') \to \rho (p_\rho) K (p_K) \right) 
&=&
\frac{2 g_{K_1 \rho K} }{m_{K_1} m_\rho} 
\Bigg[ (p'.p_\rho) ( \epsilon_\rho . \epsilon_{K_1} )
- (p'.\epsilon_\rho)(p_\rho.\epsilon_{K_1}) \Bigg] \nonumber \\ 
&=& 
\frac{ g_{K_1 \rho K} }{m_{K_1} m_\rho} 
   \Bigg[ \left( P^2 + P.Q\right)  
    ( \epsilon_\rho . \epsilon_{K_1} ) 
- \left(P.\epsilon_\rho \right) \left(Q.\epsilon_{K_1}\right)
\Bigg] \,\,\, . 
\label{eq:22}
\eeqa
This matrix element will give the decay width \cite{Roca:2003uk}:
\beq
\Gamma_{K_1} = \frac{|g_{K_1 \rho K}|^2}{2 \pi m_{K_1}^2} 
q' \bigg( 1 + {2 \over 3} \frac{q'^2}{m_\rho^2} \bigg) \,\,\, , 
\label{eq:23}
\eeq
with $q' = \frac{1}{2 m_{K_1}} \lambda^{1/2}(m_{K_1}^2,m_\rho^2,m_K^2)$, and where $p'$, $\epsilon_{K_1}$ and $p_\rho, \epsilon_\rho$ are the momentum and polarization vectors of $K_1$ and $\rho$ respectively. In the following analysis we shall neglect the masses of the leptons, the kaon and the $\rho$, where in the above we have used $p' = P$ and $p_\rho = (P + Q)/2$.  
 
\par The final 4-body decay amplitude can be written as the sum of two amplitudes:
\beq
{\cal A} = \Bigg[\frac{G_F}{\sqrt{2}} V_{tb}V_{ts}^*  
 \frac{\alpha m_b}{ \pi L^2} 
\left( \frac{ g_{K_1 \rho K} }{m_{K_1} m_\rho} \right)\Bigg] ~
\left(\epsilon_\rho\right)_\beta \left(  {\cal A}_R^\beta +
  {\cal A}_L^\beta \right) \,\,\, , 
\label{eq:24}
\eeq
where
\begin{eqnarray}
{\cal A}_R^\beta &=& (\bar \ell_R \gamma^\mu l_R)
\left( a_R g_{\mu\nu} -b_R {P_\mu L_{\nu}} +
ic_R ~\epsilon _{\mu \nu \alpha \beta} P^\alpha L^\beta \right ) 
\frac{g^{\nu\alpha}-P^\nu P^\alpha/m_{K_1}^2}
{ P^2-m_{K_1}^2 +i m_{K_1} \Gamma_{K_1}} \nonumber \\ 
&& \hspace{1.5cm} \times \Bigg[ \left(P^2 + P.Q\right) g_{\alpha \beta} - P_\beta
Q_\alpha  \Bigg] \,\,\, ,   \label{eq:25}\\
{\cal A}_L^\beta &=& (\bar \ell_L \gamma^\mu l_L)
\left( a_L g_{\mu\nu} -b_L {P_\mu L_{\nu}} +
ic_L ~\epsilon _{\mu \nu \alpha \beta} P^\alpha L^\beta \right ) 
\frac{g^{\nu\alpha}-P^\nu P^\alpha/m_{K_1}^2}
{ P^2-m_{K_1}^2 +i m_{K_1} \Gamma_{K_1}}  \nonumber \\ 
&& \hspace{1.5cm} \times \Bigg[ \left(P^2 + P.Q\right) g_{\alpha \beta} - P_\beta
Q_\alpha  \Bigg] \,\,\, . 
\label{eq:26}
\end{eqnarray}
The  $a_R$, $b_R$, $c_R$ and $a_L$, $b_L$, $c_L$ can be expressed as:
\begin{eqnarray}
a_L &=& -C_7  \left[2 (P\cdot L) g_+ + L^2 (g_+ +g_-) \right]
+\frac{(C_9^{eff} -C_{10}) f}{2m_b}L^2 \,\,\, ,
\label{eq:27}\\
b_L &=& -2C_7 ( g_+  -  L^2 h )-\frac{(C_9^{eff} -C_{10}) a_+}{m_b}L^2 \,\,\,
      ,\label{eq:28}   \\
c_L &=& -2C_7  g_+ -\frac{(C_9^{eff}-C_{10}) g}{m_b}L^2 \,\,\, ,\label{eq:29} \\
a_R &=& -C_7  \left[2 (P\cdot L) g_+ + L^2 (g_+ +g_-) \right]
+\frac{(C_9^{eff} + C_{10}) f}{2m_b}L^2 \,\,\, , \label{eq:30}\\
b_R &=& -2C_7 (  g_+ - L^2 h )-\frac{(C_9^{eff} + C_{10}) a_+}{m_b}L^2 \,\,\,
, \label{eq:31}\\ 
c_R &=& -2C_7  g_+ -\frac{(C_9^{eff} + C_{10}) g}{m_b}L^2 \,\,\, . \label{eq:32}
\end{eqnarray}

\begin{figure}[htb]
\bcen
\hskip -.5in
\epsfig{file=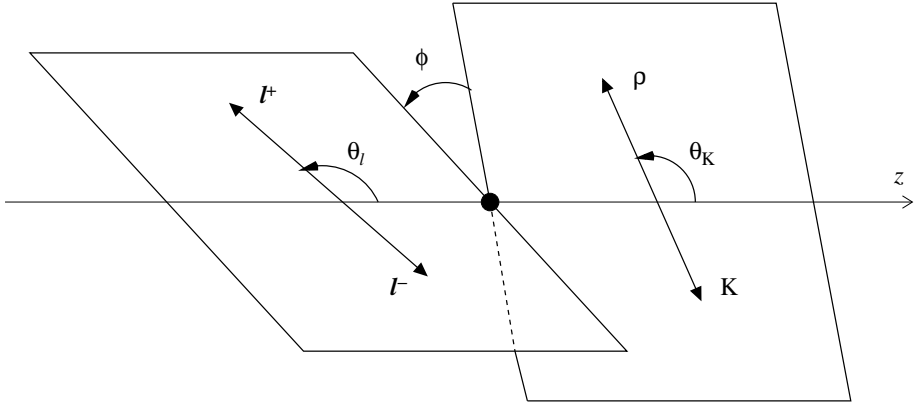,width=.7\textwidth}
\caption{\it The definition of the kinematical variables in the decay $B \to K_1(\to \rho K) \ell^+ \ell^-$.}
\label{fig:kine}
\ecen
\end{figure}

\par As such the decay rate can be computed, with the result:
\beqa
\frac{d^5 \Gamma }{dp^2dl^2 d\cos \theta_ K d\cos \theta_+ d \phi} &=&
\frac{2\sqrt{\lambda}}{128 \times 256 \pi^6 m_B^3}
\left| \frac{G_F}{\sqrt{2}} V_{tb} V_{ts}^* 
\frac{\alpha m_b}{\pi L^2} 
\left( \frac{ g_{K_1 \rho K} }{m_{K_1} m_\rho} \right) \right|^2 
\left(|A_R|^2+|A_L|^2\right) \,\,\, ,
\label{eq:33}
\eeqa
where 
\beq
|A_{\{L,R\}}|^2 = \left( - g_{\alpha \beta} +
  \frac{(p_\rho)_\alpha (p_\rho)_\beta}{m_\rho^2} \right) 
\left({\cal A}_{\{L,R\}}\right)^\alpha
\left({\cal A}_{\{L,R\}}^*\right)^\beta \,\,\, ,
\label{eq:34}
\eeq
where the various angles used above are as shown in figure \ref{fig:kine}. Recall that we shall present our results in terms of the vectors $P$, $Q$, $L$ and $N$, by use of the transformation:
$$p_\rho \to \frac{P+Q}{2} \,\,\, , \hspace{1cm} p_K \to \frac{P -
  Q}{2} \,\,\, , \hspace{1cm} p_+ \to \frac{L + N}{2} \,\,\, ,
\hspace{1cm} p_- \to \frac{L -N}{2} \,\,\, .$$

\par Using the kinematics as prescribed in Kim {\it et al.}\cite{Kim:2000dq}, that is, where we set $p = \sqrt{p_{K_1}^2}$, $l = \sqrt{(p_+ + p_-)^2}$ and $\lambda = \frac{1}{4} ( m_B^2 - p^2 - l^2)^2 - p^2 l^2$. Furthermore, we shall introduce various angles, namely $\theta_K$ as the polar angle of the $\rho$ momentum in the rest frame of the $K_1$ meson with respect to the helicity axis, {\it i.e.} the outgoing direction of $K_1$. Similarly $\theta_+$ as the polar angle of the positron in the dilepton CM frame with respect to the $K_1$ momentum. And finally $\phi$ as the azimuthal angle between these planes, that is, the $K_1 \to \rho K$ and $B \to K_1 \ell^+ \ell^-$ planes. In this case:
\beqa
|A_L|^2 &=& 
\frac{1}{(P^2 - m_{K_1}^2)^2 + (m_{K_1} \Gamma_{K_1})^2}   
{1 \over 2}
\Bigg[
|a_L|^2 \Bigg\{ 2 \left(P^2 + 2 P.Q\right) \left((L.P)^2 - (N.P)^2\right)
  + 2 \big( (N.Q)^2    \nn \\
&& - (L.Q)^2 + L^2 (P.Q)  
 - N^2 (P.Q) \big) P^2  + 4 \left( (L.Q) (L.P) + (N.P) (N.Q) \right) (P.Q)
    + (L^2 - N^2)  \nn \\
&& \times (P^2 + Q^2) P^2 \Bigg\}  +  |b_L|^2 
  \Bigg\{ \bigg(2 (L.P)^2  
 - 2 (N.P)^2 + (N^2 - L^2) P^2 \bigg)
     \bigg( (P^2 + 2 (P.Q)) (L.P)^2   \nn \\
&& - (P^2 + (P.Q))^2 L^2  
 + 2 (L.Q) (P.Q) (L.P) - (L.Q)^2 P^2    \bigg)   \Bigg\} 
 + |c_L|^2
\Bigg\{ 
\bigg[ N^2 Q^2 P^2 - 2 (N.Q)^2P^2   \nn \\
&& +  \bigg(2 P^4 + (4 (P.Q) + Q^2) P^2 + 2 (P.Q)^2 \bigg) P^2  
 + 2 (P.Q)^2\bigg] (L.P)^2  
 + 4 (N.P) (N.Q) (L.Q)  \nn \\
&& \times (L.P) P^2 
  - 2 (N.P)^2 (L.Q)^2 P^2 
 - \bigg[ -  2 \bigg(P^4 + (2 (P.Q) + Q^2) P^2 + (P.Q)^2 \bigg) (N.P)^2 \nn \\
&&   
 + 4 (N.Q) (P.Q) (N.P) P^2 
 + \bigg( - 2 P^2 (N.Q)^2 + N^2 (P^2 Q^2 - (P.Q)^2)
   + L^2 (2 P^4 + (4 (P.Q)  \nn \\ 
&& + Q^2) P^2  + (P.Q)^2) \bigg) P^2
\bigg]L^2  
\Bigg\} 
 + 4 Re(a_L b_L^*) 
\Bigg\{  - (P^2 + 2 (P.Q))(L.P)^3  - 2 (L.Q) (P.Q) (L.P)^2  \nn \\
&& +  \bigg( P^2 (L.Q)^2 +  (P^2  + (P.Q))^2 L^2 
+ (N.P) (N.Q) (P.Q) + (N.P)^2 (P^2 + 2 (P.Q)) \bigg) (L.P)  \nn \\
&& + (L.Q) (N.P) \bigg( (N.P) (P.Q) - (N.Q) P^2 \bigg) \Bigg\} 
 + 4 Re(a_L c_L)
\Bigg\{ (N.Q) (P.Q) (L.P)^2 + (N.P)  \nn \\ 
&& \times (L.Q)^2 P^2 
 - \bigg( (N.Q) P^2 + (N.P) (P.Q) \bigg) (L.Q) (L.P)
\Bigg\}  - 4 Re(b_L c_L^*) 
\Bigg\{ \bigg( (L.Q) P^2  \nn \\
&& - (L.P) (P.Q) \bigg)  
  \bigg( - (N.Q) (L.P)^2 + (L.Q)(N.P)(L.P)
      + \bigg[ (N.Q) P^2    
- (N.P)(P.Q) \bigg] L^2 \bigg) \Bigg\}  \nn \\
&&+ 4 \widetilde{(LNPQ)}
    \bigg( (L.P) (P.Q) - (L.Q) P^2 \bigg) 
\bigg( Im(a_L b_L^*) + (N.P) Im(b_L c_L^*) \bigg) \nn \\
&&  + 4 Im(a_L c_L^*) \widetilde{(LNPQ)}
    \bigg( (N.Q) P^2 - (N.P)(P.Q) \bigg)  \Bigg] \,\,\, , 
\label{eq:35}
\eeqa
\beqa
|A_R|^2 &=& 
\frac{1}{(P^2 - m_{K_1}^2)^2 + (m_{K_1} \Gamma_{K_1})^2}   
{1 \over 2}
\Bigg[
|a_R|^2 \Bigg\{ 2 \left(P^2 + 2 P.Q\right) \left((L.P)^2 - (N.P)^2\right)
  + 2 \big( (N.Q)^2    \nn \\
&& - (L.Q)^2 + L^2 (P.Q)  
 - N^2 (P.Q) \big) P^2  + 4 \left( (L.Q) (L.P) + (N.P) (N.Q) \right) (P.Q)
    + (L^2 - N^2)  \nn \\
&& \times (P^2 + Q^2) P^2 \Bigg\}  +  |b_R|^2 
  \Bigg\{ \bigg(2 (L.P)^2  
 - 2 (N.P)^2 + (N^2 - L^2) P^2 \bigg)
     \bigg( (P^2 + 2 (P.Q)) (L.P)^2   \nn \\
&& - (P^2 + (P.Q))^2 L^2  
 + 2 (L.Q) (P.Q) (L.P) - (L.Q)^2 P^2    \bigg)   \Bigg\} 
 + |c_R|^2
\Bigg\{ 
\bigg[ N^2 Q^2 P^2 - 2 (N.Q)^2P^2   \nn \\
&& +  \bigg(2 P^4 + (4 (P.Q) + Q^2) P^2 + 2 (P.Q)^2 \bigg) P^2  
 + 2 (P.Q)^2\bigg] (L.P)^2  
 + 4 (N.P) (N.Q) (L.Q)  \nn \\
&& \times (L.P) P^2 
  - 2 (N.P)^2 (L.Q)^2 P^2 
 - \bigg[ -  2 \bigg(P^4 + (2 (P.Q) + Q^2) P^2 + (P.Q)^2 \bigg) (N.P)^2 \nn \\
&&   
 + 4 (N.Q) (P.Q) (N.P) P^2 
 + \bigg( - 2 P^2 (N.Q)^2 + N^2 (P^2 Q^2 - (P.Q)^2)
   + L^2 (2 P^4 + (4 (P.Q)  \nn \\ 
&& + Q^2) P^2  + (P.Q)^2) \bigg) P^2
\bigg]L^2  
\Bigg\} 
 + 4 Re(a_R b_R^*) 
\Bigg\{  - (P^2 + 2 (P.Q))(L.P)^3  - 2 (L.Q) (P.Q) (L.P)^2  \nn \\
&& +  \bigg( P^2 (L.Q)^2 +  (P^2  + (P.Q))^2 L^2 
+ (N.P) (N.Q) (P.Q) + (N.P)^2 (P^2 + 2 (P.Q)) \bigg) (L.P)  \nn \\
&& + (L.Q) (N.P) \bigg( (N.P) (P.Q) - (N.Q) P^2 \bigg) \Bigg\} 
 - 4 Re(a_R c_R)
\Bigg\{ (N.Q) (P.Q) (L.P)^2 + (N.P)  \nn \\ 
&& \times (L.Q)^2 P^2 
 - \bigg( (N.Q) P^2 + (N.P) (P.Q) \bigg) (L.Q) (L.P)
\Bigg\} + 4 Re(b_R c_R^*) 
\Bigg\{ \bigg( (L.Q) P^2  \nn \\
&& - (L.P) (P.Q) \bigg)  
  \bigg( - (N.Q) (L.P)^2 + (L.Q)(N.P)(L.P)
      + \bigg[ (N.Q) P^2    
- (N.P)(P.Q) \bigg] L^2 \bigg) \Bigg\}  \nn \\
&&+ 4 \widetilde{(LNPQ)}
    \bigg( (L.P) (P.Q) - (L.Q) P^2 \bigg) 
\bigg( - Im(a_R b_R^*) + (N.P) Im(b_R c_R^*) \bigg) \nn \\
&&  + 4 Im(a_R c_R^*) \widetilde{(LNPQ)}
    \bigg( (N.Q) P^2 - (N.P)(P.Q) \bigg)  \Bigg] \,\,\, ,  
\label{eq:36}
\eeqa
where $\widetilde{(ABCD)} = \varepsilon_{\alpha \beta \gamma \delta} A^\alpha B^\beta C^\gamma D^\delta$. 

\par The $p^2$ integration is performed using the narrow width approximation of the $K_1$ decay, {\it i.e.}:
\beq
\lim_{\Gamma_{K_1} \to 0} \frac{m_{K_1} \Gamma_{K_1}}{ (P^2 -
  m_{K_1}^2)^2 + (m_{K_1} \Gamma_{K_1})^2} = \pi \delta(P^2 - m_{K_1}^2) \,\,\, . 
\eeq
As such, the total decay width can be expressed as:
\beqa
\Gamma &=&
\int p^2 \delta(p^2 - m_{K_1}^2) 
\int_{4 m_\ell^2}^{(m_B - m_K1)^2} l^2 
\int_{-1}^1 d(cos\theta_K)
\int_{-1}^1 d(cos\theta_+)
\int_0^{2\pi} d\phi
\frac{2\sqrt{\lambda}}{128 \times 256 \pi^6 m_B^3}   \nn  \\
&& \hspace{1cm} \times \left| \frac{G_F}{\sqrt{2}} V_{tb} V_{ts}^* 
\frac{\alpha m_b}{\pi L^2} \right|^2 
\frac{2 \pi^2}{m_{K_1} m_\rho^2 q' \left(1 + {2 \over 3}
    \frac{q'^2}{m_\rho^2} \right)} 
(|A_R|^2+|A_L|^2) \,\,\, ,
\eeqa
with 
$q' = \frac{1}{2 m_{K_1}} \lambda^{1/2}(m_{K_1}^2,m_\rho^2,m_K^2)$ and $\lambda = \frac{1}{4} ( m_B^2 - p^2 - l^2)^2 - p^2 l^2$. 


\section{Results \label{section:4}}

\par The form factors for the radiative mode $B \to K_1(1270) \gamma$,
as given by Cheng and Chua \cite{Cheng:2004yj}, assumed that the
physical states $K_1(1270)$ and $K_1(1400)$ were mixtures of the
angular momentum eigenstates $K_{1A}$ and $K_{1B}$, where the mixing
angle between these states is not known precisely (though it is
believed to be such as to cause the maximal mixing between the
states). Where the hypothesis of mixing between the two states
naturally 
explains the suppression of one of the decay modes with respect to the
other. In reference \cite{Cheng:2004yj} the mixing angles suggested
were $\theta = \pm 37^\circ, \pm 58^\circ$. The negative values of the
mixing angles suggest the suppression of $B \to K_1(1270) \gamma$ as
compared to $B \to K_1(1400) \gamma$, which is disfavoured (although
not conclusively ruled out) from the observation of the radiative
decay mode $B \to K_1(1270,1400) \gamma$ by the BELLE collaboration
\cite{Yang:2004as}.  

\par Although the prescription of mixing between the states helps to
explain the suppression of one of the modes, as compared to the other,
the form factors as given in reference \cite{Cheng:2004yj} predict a
lower value of the branching ratio for $B \to K_1(1270) \gamma$ as
compared to experimental results. Note that there have been many attempts in
references \cite{Kwon:2004ri} to address this issue, where these attempts
essentially predict a much larger value of the zero recoil value of
the form factors. For our analysis we have used the form factors as
given by Cheng and Chua \cite{Cheng:2004yj}. Our analytical results
for the LEET form factors and the differential decay rate retains the
same form for any possible increase in the zero recoil value of the
form factors. For our analysis we have used the mixing angle between
the two angular momentum eigenstates ($K_{1A}, K_{1B}$) to be ($\theta
=$) $58^\circ$. Our SM value of the branching ratio for $\btokrhokll$,
using the input parameters as defined in Appendix \ref{appendix:b}, is
$2.3 \times 10^{-7}$. 

\begin{figure}[htb]
\bcen
\hskip -.5in
\epsfig{file=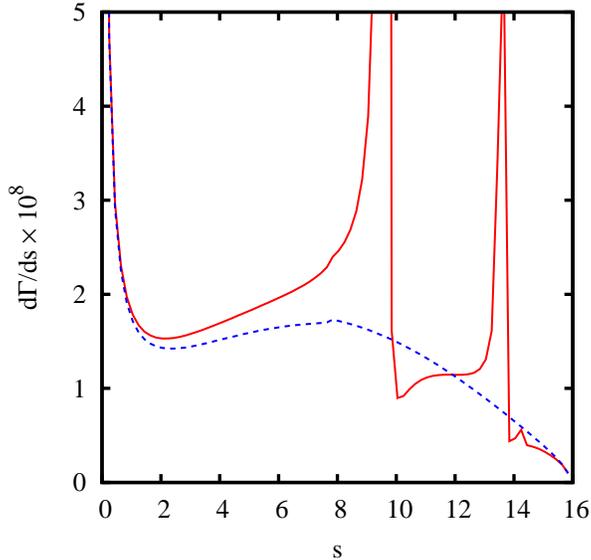}
\caption{{\it The differential branching ratio as a function of
    dilepton invariant mass. The two lines correspond to the result of
    switching on (solid line) and off (dashed line) the charmomium
    resonances in $C_9^{eff}$.}}  
\label{fig:1}
\ecen
\end{figure}

\par In figure \ref{fig:1} we have shown the variation of the differential
branching ratio of $\btokrhokll$ as a function of the dileptonic
invariant mass. The two different lines in the plot correspond to the
results of including and excluding the long distance charm resonances,
where these long-distance effects can be included with the
redefinition of $C_9^{eff}$. For this purpose we have used the
prescription as given in Kruger and Sehgal \cite{Kruger:1996cv}.  
Note that use of the LEET for obtaining the vector and axial vector form factors
is justified more for relatively large values of $s$, where at the moment
there are no first principle determinations of these form factors 
similar to the ones determined for the tensor form factors by Cheng
and Chua \cite{Cheng:2004yj}. We therefore continue using the LEET
based form 
factors for the entire range. As and when more accurate values are
available for the low $s$-region our expressions can easily be
reevaluated for a fresh plot. 
 
\par
\begin{figure}[htb]
\bcen
\epsfig{file=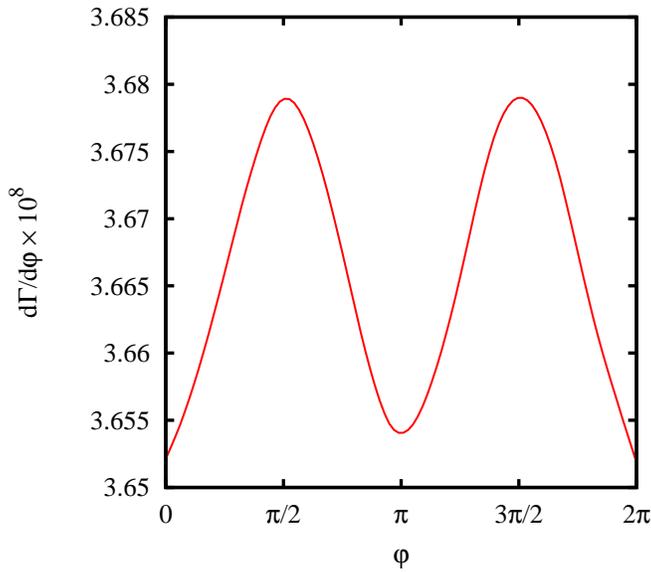}
\caption{{\it The differential branching ratio as a function of the azimuthal angle $\phi$.}}  
\label{fig:2}
\ecen
\end{figure}

 In figure \ref{fig:2} we have shown the plot of the differential
branching ratio as a function of the azimuthal angle between the
planes $\rho K$ and $\ell^+ \ell^-$. And in our final plot we have shown
the dependence of the differential branching ratio as a function of
the scattering angle $\theta_\ell$ in $\ell^+ \ell^-$, as defined in
figure \ref{fig:kine}. 

\par We should like to note, at this point, that for any new physics model that can be effectively absorbed by the ``standard'' set of Wilson coefficients in the effective Hamiltonian,
our analytic results given in previous sections can be used to obtain the corresponding change in the angular correlations. 

\par As one final remark we may integrate our differential decay rate
over the final state hadrons to get the decay rate of the process $B
\rightarrow K_1 \ell^+ \ell^-$. Note that by using our defintion of
the form factors we can relate these to the ones for $K^*(890)$, under
the substitutions $ V \leftrightarrow - A$ and $T \leftrightarrow
T_5$, such that the corresponding decay rates are obtained easily (by
means of this substitution). It is easily seen then that the location
of the zero in the FB asymmetry of this integrated decay rate is the
same as in the $K^*(890)$ case, with the numerical value of the form
factors being different. However, though the zeroes can be related,
the overall shape of the FB asymmetries could be different from $B \to
K^*(890)$.  

\par To conclude, the mode which we have studied above, within the SM level, can in
principal be measured at present $B$-factories. Various angular
correlations of this mode can also be studied in future SuperB
factories. The study of the various angular correlations in
$\btokrhokll$ can provide us with a very useful cross-checking tool
for the SM and possible new physics in $b \to s \ell^+ \ell^-$
transitions. In this pursuit we have given the LEET form factors for
$B \to K_1(1270) \ell^+ \ell^-$, which could be useful in not only
testing the operator structure of the SM but also the existence of
possible new physics beyond it.  

\begin{figure}[h]
\bcen
\epsfig{file=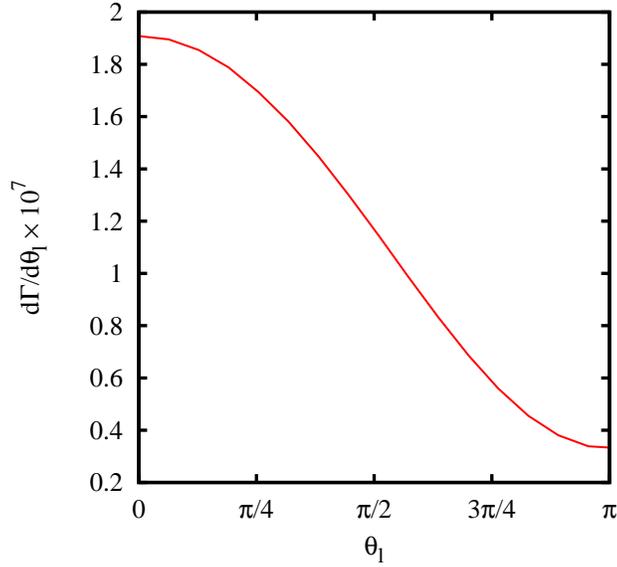}
\caption{{\it The differential branching ratio as a function of $\theta_\ell$.}} 
\label{fig:3}
\ecen
\end{figure}


\section*{Acknowledgments}

The work of SRC is supported by the Department of Science \&
Technology (DST), India.  SRC would also like to thank the Theory Division, KEK, for their local hospitality, where the part of this work was done. The work of NG
is supported by JSPS under the grant no. P06043.


\appendix


\section{The form factors \label{appendix:a}}

\noindent Firstly, it is important to note that the additional
negative sign in equation (\ref{eq:19}) is due to the difference in
our definition of $\varepsilon$, as compared to that of Cheng and Chua
\cite{Cheng:2004yj}, Charles {\it et al.}\cite{Charles:1998dr} and Kim
{\it et al.}\cite{Kim:2000dq}. Furthermore, note that the definitions
of Cheng {\it et al.} are the same as that of Charles {\it et al.},
but differ from that of Kim {\it et al.} by a sign, which can be taken
into account by changing the sign in equation (\ref{eq:19}).   

\par The form factors, as defined in Cheng {\it et al.} (for all the form factors except $Y_{B3}$)\cite{Cheng:2004yj}, are:  
\beq
F(q^2) = \frac{F(0)}{1 - a (q^2/m_B^2) + b (q^2/m_B^2)^2} \,\,\,
. \label{eq:app:1} 
\eeq
For $Y_{B3}$ we use \cite{Cheng:2004yj}:
\beq
F(q^2) = \frac{F(0)}{\left(1 - q^2/m_B^2\right) 
\left(1 - a (q^2/m_B^2) + b (q^2/m_B^2)^2 \right)} \,\,\, . \label{eq:ap:2}
\eeq
The numerical values of the factors appearing in equation (\ref{eq:22}) and equation (\ref{eq:23}) are given in Table \ref{table:1}.  

\begin{table}[ht]
\begin{center}
\begin{tabular}{c c c c}  \hline \hline
F         &   F(0)  &   a    &   b  \\  \hline
$Y_{A1}$  &   0.11  &  0.68  &  0.35  \\
$Y_{A3}$  &   0.19  &  1.02  &  0.35  \\
$Y_{B1}$  &   0.13  &  1.94  &  1.53  \\
$Y_{B3}$  & - 0.07  &  1.93  &  2.33  \\
\hline \hline
\end{tabular}
\caption{{\it The form factors for the $Y_{A_i}$ and $Y_{B_i}$}\cite{Cheng:2004yj}.}
\label{table:1}
\end{center}
\end{table}


\section{Input Parameters \label{appendix:b}}

\bcen
$m_B = 5.26$ GeV, ~~$m_{K_1} = 1.27$ GeV, ~~ $m_b = 4.8$ GeV, ~~ $m_\rho =
0.77 $ GeV, ~~ $m_K = 0.134$ GeV, \\
$\alpha = \frac{1}{129}$, ~~ $G_F = 1.17 \times 10^{-5}$ GeV$^{-2}$,
~~ $|V_{tb} V_{ts}^*| = 0.0385$, 
\\
$C_7 = - 0.3, ~~ \mathrm{and} ~~ C_{10} = - 4.6,$
\ecen


\end{document}